\documentclass[letterpaper,12pt]{article}
\usepackage{bejournal}          
\usepackage{graphicx,amsmath,verbatim,times,url}
\usepackage[authoryear,sectionbib]{natbib} 
\usepackage[usenames,dvipsnames]{color}

\renewcommand{\vec}[1]{\boldsymbol{#1}}
\newcommand{\N}{\text{N}}
\newcommand{\Ga}{\text{Ga}}
\renewcommand{\d}{\phantom{2}}
\title{Bayesian identification of protein differential expression in
  multi-group isobaric labelled mass spectrometry data}
\author{Howsun Jow, Richard J. Boys and Darren J. Wilkinson%
\thanks{School of Mathematics \& Statistics, Newcastle University, U.K.}}
 
\begin{document}

\originalmaketitle
\begin{abstract}
  In this paper we develop a Bayesian statistical inference approach
  to the unified analysis of isobaric labelled MS/MS proteomic data
  across multiple experiments. An explicit probabilistic model of the
  log-intensity of the isobaric labels' reporter ions across multiple
  pre-defined groups and experiments is developed. This is then used
  to develop a full Bayesian statistical methodology for the
  identification of differentially expressed proteins, with respect to
  a control group, across multiple groups and experiments. This
  methodology is implemented and then evaluated on simulated data and
  on two model experimental datasets (for which the differentially
  expressed proteins are known) that use a TMT labelling protocol.

\bigskip
\noindent 
\textbf{Keywords}: proteomics, Bayesian, mass spectrometry, MS/MS,
differential protein expression
\end{abstract}

\section{Introduction}

In recent years, isobaric molecular labelling techniques have been
developed which, in conjunction with tandem mass spectrometry (also
known as MS/MS), can be used to perform quantitative analyses of
complex protein mixtures \citep{Thompson2003, Ross2004}.  This
procedure involves using enzymes such as trypsin to digest the
proteins in the samples into their constituent peptides. Isobaric
molecular labels (one label per sample) are then chemically attached
to the resulting peptide components of the enzyme digested proteins.
The samples are then mixed and passed through a liquid chromatograph
into a mass spectrometer where tandem mass spectrometry is performed.
The relative intensities of these labels for the resulting MS/MS
spectra can be determined from the intensities of the isobaric labels'
``reporter'' ions. The MS/MS spectra are those of the constituent
peptides and these can be identified either by hand or, more usually,
using a protein sequence database. If the identified peptides are
prototyped for a protein, i.e. peptides that can be uniquely derived
from the digestion of a protein, then the relative amounts of this
protein in the labelled samples can be quantified.

The experimental process described above is inherently stochastic as
even technically identical replicates will produce different data
regardless of the accuracy of the mass spectrometer. Also there is a
wide ``dynamic'' range in the intensity of the detected peptides.
This matters because only a limited number of ions can be subject to
MS/MS.  The highest intensity ions are the ones which tend to be
selected. As such, the more abundant proteins, especially large
proteins, will tend to drown out less abundant smaller proteins.  A
variety of techniques exist by which the complex protein mixture can
be split up into simpler fractions but these do not completely
eliminate the problem.  An additional problem is that there are a
limited number of isobaric tags available. The maximum currently
commercially available is eight \citep{Choe2007}. This limits the
number of samples that can be analyzed in a single experiment.
However, it should be possible to analyze a larger number of samples
using multiple experiments that, for example, share a common reference
sample. This requires that any analysis technique used be able to link
data from multiple experiments of this type.

Apart from the peptide identification problem
\citep{befekadu2009bayesian}, the main statistical problem in
analysing MS/MS data is how to detect which of the proteins in a
complex mixture are present in significantly different amounts in a
set of isobaric labelled samples from two or more pre-determined
groups such as a control group and a set of treatment groups.
Currently standard software such as MASCOT \citep{MASCOT} is quite
limited as they attempt to detect differences between two groups
either by a informal thresholding on the fold-ratio or by using a
t-test, which in turn can be generalised to a comparison of more
groups via a one-way ANOVA.  The t-test method typically makes the
reasonable assumption that the intensities of the reporter ions for
MS/MS spectra of proteotypic peptides are log-normally distributed
\citep{Boehm2007}.  However, using the t-test approach has several
drawbacks. The main problem arises from having to test a large number
of hypotheses to determine significant differences in expression
levels between groups, as a test is needed for each of the hundreds of
proteins.  Of course, such problems can be somewhat mitigated by using
standard multiple hypotheses corrections \citep{Sidak1968,Holm1979,
  Benjamini1995}.  Another problem with using t-test is that many
suffer from low power due to proteins being quantified on the basis of
only a small number of MS/MS spectra.

To circumvent these issues, various authors have developed methods
using a more detailed model for the peptide intensity and use ANOVA
analytic techniques to determine differences between protein
expression levels for different groups and their associated
statistical significance. \citet{Keshamouni2006} describe a very
simple ANOVA model for normalised log-ratios in a single experiment
comparing a single treatment against a control. A more sophisticated
ANOVA model for multiple treatments and experiments is described in
\citet{Oberg2008}. Although their full model is somewhat
overparameterised, it is easily reduced to an identifiable model by
combining some parameters and setting others to zero. This reduced
model is very closely related to the model described in this paper.
The main innovations introduced here are that the model is fit jointly
to all of the data simultaneously, rather than using a stepwise
regression approach, that the fitting is done in a fully Bayesian way,
and that the Bayesian version of the model includes variable selection
indicators for differential expression, allowing direct inference to
be made for the probability of differential expression associated with
each protein. Note that related methods for LC--MS data are often
tailored to specific features of that data. For example, the models of
\cite{karpievitch2009statistical} and \cite{wang2012hybrid} are
tailored to a specific kind of informative missingness in the data
which does not arise for isobaric labelled MS/MS data. In the MS/MS
framework considered here, missingness is generally much less of an
issue than for LC--MS data \citep{Oberg2012}, and there is little
evidence to suggest that missing values in MS/MS data sets are
strongly informative.  Nevertheless, as for all proteomics
technologies, there is still an issue with certain peptides not being
detected at all during the first phase of the MS/MS procedure.  Our
model deals correctly with such missingness, in a fairly efficient
way.

In this paper we fit a detailed random effects model with an ANOVA
form to analyse protein expression levels, but remove or consolidate
many of the parameters that \citet{Oberg2008} find to be typically
confounded. The model is a hierarchical model which borrows strength
appropriately across multiple peptides, proteins, samples and
experiments. Unlike the stepwise regression approach described by
\citet{Oberg2008}, our fitting approach does not force each protein to
have an independent variance parameter, and correctly propagates
normalisation uncertainty without assuming approximate orthogonality
of model components. In common with \citet{Oberg2008}, our model
allows for the analysis of multiple experiments with common reference
samples.  Additionally, our method adopts a fully Bayesian approach
and therefore has all the advantages of interpretability and being
able to include prior information (where available). The Bayesian
approach also has several advantages specific to the context of
complex hierarchical models. The framework is flexible, and allows
convenient checking for over-parametrisation or variable confounding
by prior to posterior comparisons and multi-dimensional analysis of
the posterior distribution.  Importantly the model has a variable
selection form which ensures that we fit an appropriate model for the
combination of differentially expressed and non-differentially
expressed proteins.  Models without this structure have the drawback
that they inflate the error variance in the ``null'' model due to
contamination by outlying differentially expressed proteins which then
hinders the detection of differential expression.

In the next section we give more details of the experimental framework
and describe the model together with the prior-to-posterior analysis.
In section~\ref{sec:sim} we demonstrate the potential of the model to
pool information across multiple MS/MS experiments by using simulated
data. In sections~\ref{sec:plasma} and~\ref{sec:proteoRed} we analyse
two real MS/MS datasets.  The real datasets have a simple structure
but serve to demonstrate that our model captures the behaviour of real
experimental data. The dataset in section~\ref{sec:plasma} has two
technically identical replicates and so contains a negative control,
that is, there should be no differentially expressed proteins. The
dataset in section~\ref{sec:proteoRed} is provided by the ProteoRed
consortium as part of an assessment of various quantitative proteomics
methods. It contains known differentially expressed proteins within an
otherwise set of technically identical replicates. The paper concludes
in section~6 with a discussion.

\section{Methods}
\label{sec:methods}

\subsection{Experimental framework}

The experimental framework of an isobaric labelled MS/MS experiment is
as follows: the protein samples, each containing roughly equal amounts
of total protein by mass, are digested using enzymes with typically
high specificity such as trypsin i.e. they cleave the protein
molecules into peptide fragments at predictable amino acid residues.
The digested samples are then each chemically labelled with different
isobaric labelling molecules. The samples are then mixed together and
run through a liquid chromatograph into a mass spectrometer.

A selection of high intensity ions from the resulting MS spectra are
then shattered, using high energy collisions, to yield further ion
spectra called MS/MS spectra. The chemically attached molecules leave
signature or reporter ions in the MS/MS spectra. The relative
intensity of these reporter ions will give us relative quantitative
information about the amounts of the peptide fragments, corresponding
to each MS/MS spectra, in each sample.

The above constitutes a single experiment. Multiple experiments ($E$)
can of course be performed either with technical or biological sample
replicates.  The samples being analyzed are typically assigned to
pre-determined biological groups. For simplicity, only the MS/MS
spectra corresponding to peptide fragments uniquely derived from a
single protein are used in the quantitative analysis. This means that
a single MS/MS spectrum gives quantitation information about a single
protein.  Thus we need a model that will simultaneously model the
observed expression levels for $P$ proteins for the $G$ biological
groups to which the samples being analyzed are assigned. One of the
biological groups is used as the control group. Differences in protein
expression level are then measured relative to this control. The total
protein content of each sample that ends up being labelled are only
roughly equal. In our model we include a parameter which allows for
differing amounts of total labelled samples in an experiment.  In
addition, when performing the analysis across multiple experiments it
is important to be able to ``link'' the experiments. In order to
facilitate this a reference sample is selected in each experiment.
Ideally this should be a technical replicate of some standard sample
that is included in all our experiments.

\begin{figure}[th]
\includegraphics[width=\textwidth]{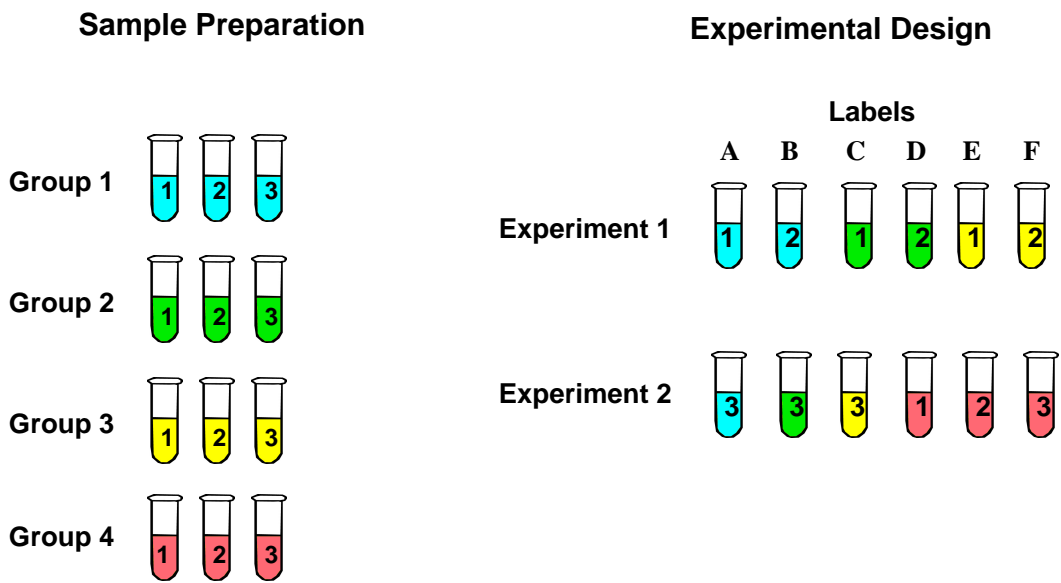}
\caption{Schematic of an experimental design for twelve samples from
  four groups (three replicate samples per group) using six available
  isobaric labels (A-F).}
\label{fig:design}
\end{figure}

An example of an experimental design is given in
Figure~\ref{fig:design}. It shows a scenario with two experiments for
twelve samples from four groups with three replicate samples per group
assuming that only six isobaric labels are available for any single
experiment.
\subsection{The Model}
\label{sec:model}
We consider a series of isobaric labelled mass spectrometric
experiments designed using $n_I$ distinctly identifiable isobaric tags
(we will typically have $n_I=6$). We assume that for each
experiment~$e$, we have (log) intensity measurements $y_{egjki}$ in
treatment group~$g$ arranged by sample/replicate number~$i$ from an
MS/MS spectrum~$k$ identified as being that of a peptide derived from
a protein~$j$. Specific peptides are therefore identified by a
particular $(j,k)$ combination. The isobaric tags are used to identify
the $(g,i)$ combinations within an experiment. The model has an ANOVA
style, starting with a sample-specific normalising constant,
$\kappa_{egi}$, and then decomposing the remaining uncertainty using a
variable selection approach in the style of \cite{Kuo98}; see
\cite{Hara2009} for a review of other possible approaches. The
benefits of Bayesian hierarchical modelling in the context of gene
expression microarrays is already well established
\citep{hein2005bgx}. Explicit inclusion of the normalisation constant
in the model ensures that important information in the data is not
lost during the normalisation process
\citep{hein2005bgx,karpievitch2012normalization,callister2006,Hill2008}.
The benefits of including the normalising constant directly in
ANOVA--style fixed and random effects models for proteomics data is
most clearly articulated in \citet{Oberg2008}, to which the reader is
referred for further details.

We model the log intensities as
\begin{equation}
y_{egjki}=\kappa_{egi}+\alpha_{jk}+\beta_{gj}\gamma_{gj}+\epsilon_{egjki},
\label{eq:model}
\end{equation}
$e=1,\ldots,E$, $g=1,\ldots,G$, $j=1,\ldots,P$,
$k=1,\dots,m_j$ and $i=1,\ldots,n_{eg}$, where, for
identifiability, the parameters are subject to the constraints
\[
\beta_{1j}=\gamma_{1j}=0,\quad j=1,\ldots,P\quad\text{and}\quad 
\kappa_{eg_e1}=0,\quad e=1,\ldots,E,
\label{eq:constraints}
\]
where $g_e$ is the group in experiment~$e$ whose first sample is used
as the reference sample. In general, differential expression will be
measured against a control group and this group will be labelled
group~1. The control group will commonly also provide a reference
sample in each experiment (giving $g_e=1$), though sometimes the
experimenter might choose different groups to provide the reference
samples. Thus the number of isobaric labels used in each experiment is
$\sum_g n_{eg}=n_I$. Note that a MS/MS spectrum is assumed to be
``assigned'' to one and only one protein.

In a typical experiment some reporter ions do not show up in a
particular MS/MS spectrum. This corresponds to missing data in the
model and can be dealt with by treating these unobserved values in the
usual Bayesian MCMC way, that is, by including the missing data in the
model as unobserved parameters and stochastically imputing them along
with all other unknowns. This leads to the correct posterior
distribution (which is less concentrated than it would have been in
the case of full observation), but at the expense of poorer mixing of
the MCMC scheme. However, a more efficient implementation is possible
by using a ``ragged array" with varying dimension indices, which
avoids creating variables corresponding to the missing data. We adopt
this latter scheme in the implementation we describe in
section~\ref{sec:postspec}, as it has significantly better mixing
properties than the naive approach.

The interpretation of the parameters is as follows. The parameters
$\kappa_{egi}$ are normalization constants for sample $i$ in group $g$
of experiment $e$ relative to the reference sample in that experiment,
that is, the log-ratio of the total amount of sample $i$ (in group $g$
of experiment $e$) with respect to the reference sample (sample~1 in
group~$g_e$ of experiment~$e$). The $\alpha_{jk}$ parameters are the
mean log-intensities of a peptide $k$ (corresponding to a reporter ion
in an MS/MS spectrum) assigned to protein~$j$ for labelled samples
from (control) group~1.  The $\beta_{gj}$ parameters are binary
indicating whether or not the protein~$j$ is differentially expressed
for biological group~$g$ with respect to (control) group~1 and the
$\gamma_{gj}$ parameters are the difference in the mean log-expression
level of protein~$j$ between group~$g$ and (control) group~1. Finally,
the $\epsilon_{egjki}$ are independent and identically distributed
noise terms following a normal $\N(0,\sigma^2)$ distribution with zero
mean and variance~$\sigma^2$.

\subsection{Bayesian Inference}

The inference task is to make statistically valid statements about the
unknown model parameters
$(\vec{\kappa},\vec{\alpha},\vec{\beta},\vec{\gamma},\sigma)$ that
describe the unknown normalisation factors, the mean expression levels
in the different groups and the experimental noise. The Bayesian
statistical inference approach combines information from the
data~$D=\{y_{egjki}\}$ with that from prior information using Bayes
Theorem, and describes this through the posterior distribution.  If we
assume that the prior distribution for each model parameter can be
specified independently, the posterior distribution is
\begin{align*}
\pi(\vec{\kappa},\vec{\alpha},\vec{\beta},\vec{\gamma},\sigma|D)\propto
&\prod_{e,(g,i)\neq(g_e,1)}\pi(\kappa_{egi}) \times \prod_{jk}\pi(\alpha_{jk}) \times  
\prod_{g\neq 1,j}\pi(\beta_{gj})\pi(\gamma_{gj}) \times \pi(\sigma)  \\
&\quad \times \sigma^{-n}
\exp\left\{-\frac{1}{2\sigma^2}\sum_{egjki} \left(y_{egjki} - \kappa_{egi} - \alpha_{jk} - \beta_{gj}\gamma_{gj}\right)^2\right\},
\end{align*}
where $n=En_I\sum_j m_j$ is the total number of log-intensity
measurements for the isobaric labels and $\pi(\cdot)$ denote prior
probability (density) functions.

\subsubsection{Prior Distribution on Model Parameters}

\label{sec:priorspec}

\begin{figure}[t]
\includegraphics[width=\textwidth]{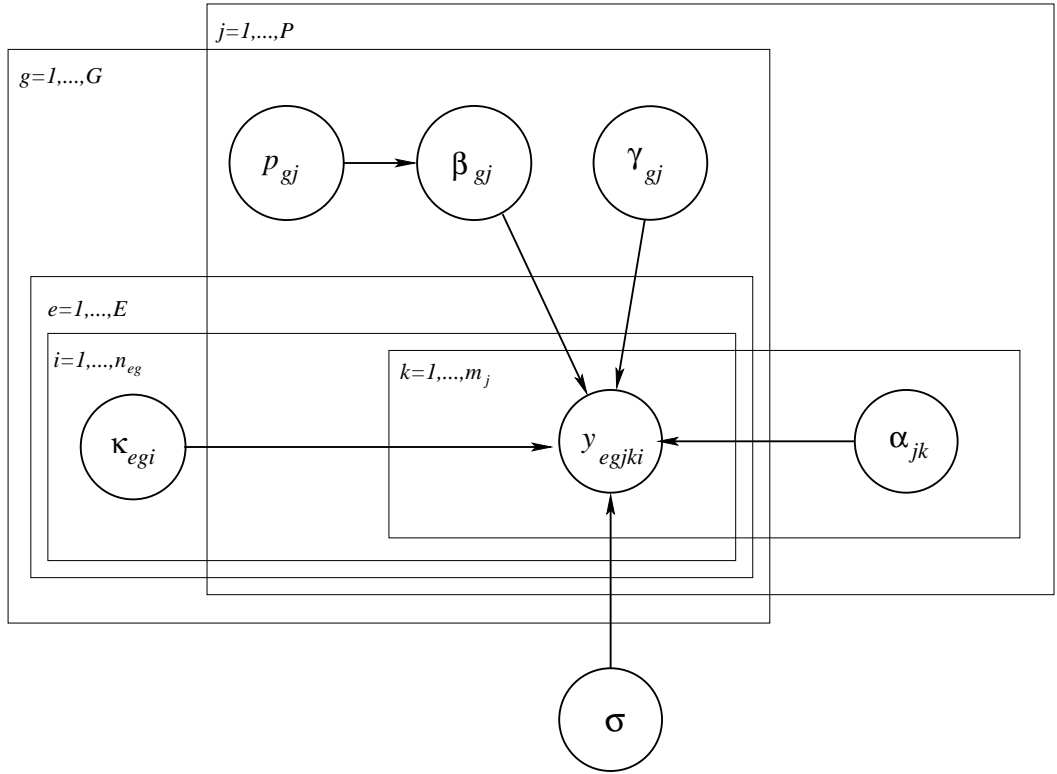}
\caption{Plate diagram for the essential model structure. The directed acyclic graph (DAG) over the random variables represents the factorisation of the joint distribution of unobserved and observed model variables, and is used to construct posterior inference schemes. Fixed hyperparameters are omitted for clarity.}
\label{fig:model}
\end{figure}

The Bayesian approach allows for additional information to be
incorporated by using a prior distribution on all model parameters
$(\vec{\kappa}, \vec{\alpha},\vec{\beta},\vec{\gamma},\sigma)$.
Although each analyst should incorporate their own prior beliefs into
the prior distribution, we have chosen to represent prior beliefs by
using standard distributions. This leaves the analyst to decide on
their own choice for the parameters of these distributions. The prior
distributions are, for $e=1,\ldots,E$, $g=1,\ldots,G$, $j=1,\ldots,P$,
$k=1,\dots,m_j$ and $i=1,\ldots,n_{eg}$
\begin{align*}
\kappa_{egi} &\sim \N(a_\kappa, 1/b_\kappa),&
\alpha_{jk}&\sim \N(a_\alpha,1/b_\alpha),\\
\beta_{gj}|p_{gj}&\sim \text{Bern}(p_{gj}),&
p_{gj}&\sim \text{Beta}(a_p,b_p),\\
\gamma_{gj}&\sim \N(a_\gamma,1/b_\gamma),&
\sigma^{-2}&\sim\Ga(a_\sigma,b_\sigma),
\end{align*}
where $\Ga(a,b)$ is a gamma distribution with mean $a/b$,
$\text{Bern}(p)$ is a Bernoulli distribution with success
probability~$p$ and $\text{Beta}(a,b)$ is a Beta distribution with
mean $a/(a+b)$. The essential model structure over the unknowns is
displayed as a plate diagram in Figure~\ref{fig:model}.

We suggest the following default choices for the prior parameters.  In
an ideal series of experiments, the experimenter uses the same total
amount of reporter in each experiment. In other words, experimenters
try to design their experiments so that the $\kappa_{egi}$ are close
to zero.  We input this information by taking $a_\kappa=0$ and
$b_\kappa=1/9$.  In our experience, log-intensities for reporter ions
typically range between 4 and 16 and so taking this range as four
standard deviations gives $a_\alpha=10$ and $b_\alpha=1/9$.  In most
analyses of this kind, the proportion of differentially expressed
proteins will be low, say around 5\%. It is possible to undertake the
analysis with this proportion fixed at such a user-specified value.
However, as results may be very sensitive to the value chosen, we
prefer to put a prior distribution on the proportion and learn about
it from the data. We suggest taking $a_p=1$ and $b_p=19$.

Naturally the degree of differential expression in
proteins (relative to the control group) will vary but we anticipate a
typical fold-change to be around one (zero on the log-scale) and so
recommend taking $a_\gamma=0$ and $b_\gamma=1$.   The level of measurement accuracy of the reporter
ion log-intensity measurements is particularly difficult to assess.
Therefore we suggest taking a quite weak prior distribution
for~$\sigma$ by using $a_\sigma=b_\sigma=1/1000$.  Overall we regard
these choices of prior parameters as inputting relatively weak prior
information. Some analysts may feel they have stronger views to
include in their prior distribution and so may feel justified in
taking quite different choices to these ``default'' ones.  We will use
our default choices in the subsequent analyses we present in this
paper.

\subsubsection{Posterior analysis}
\label{sec:postspec}
Unfortunately, the posterior distribution is analytically intractable
for any significant number of proteins and peptides. However, it is
possible to simulate realizations from this distribution by using
computer-intensive Markov chain Monte Carlo (MCMC) techniques; see
\citet{GamermanL06}. Essentially this algorithm simulates a Markov
chain which has the posterior distribution as its equilibrium
distribution.  Thus, after the algorithm has converged (in
distribution), all subsequent realizations will have the required
(posterior) distribution. We have chosen to implement our model using
the free open source software JAGS \citep{Plummer2003}.  The JAGS code
for our model and the key steps in this Gibbs sampling scheme are
given in the appendix.  Note that the indexing used in this code is
slightly different to that in~\eqref{eq:model} to deal with missing
data more efficiently: it uses a ``ragged array" structure to avoid
creating nodes for missing values. We have found that this fairly
simple one-variable-at-a-time Gibbs updating scheme works very
effectively for this class of models and so more sophisticated MCMC
schemes have not been pursued. We have also developed a full R package
implementing the model and this is available as the R-Forge package
\textbf{dpeaqms} \citep{dpeaqms}.  In more challenging scenarios,
single variable updating schemes for variable selection problems can
be inefficient, and some authors have proposed methods (such as
collasping and then joint updating) which can improve mixing in this
case \citep{Bottolo10,Davies14}.  However, such methods are not
integrated into off-the-shelf MCMC engines such as JAGS.
 
\section{Simulated Data Analysis}
\label{sec:sim}
In order to evaluate the performance of the statistical inference
technique outlined previously, the methods were tested on data
simulated from the model $\eqref{eq:model}$. The experimental scenario
we consider is one in which there are $G=4$ groups consisting of a
control group (CTL) and three treatment groups (TRT1, TRT2 and TRT3)
and we have $n_I=6$ isobaric tags to use. We assume that differential
expression is determined with respect to the control group and so
label this as group $g=1$; the other groups TRT1, TRT2 and TRT3 we
label as $g=2,3,4$ respectively. The simulated dataset we constructed
consists of $12$ labelled samples and, as this must have resulted from
more than one experiment, we assume that there were $E=2$ experiments,
with six tagged samples per experiment. We also assumed that overall
there were three samples from each group. The samples were assigned to
the six tag reporter ions according to the pattern (CTL, CTL, TRT1,
TRT1, TRT2, TRT2) for the first experiment and (CTL, TRT1, TRT2, TRT3,
TRT3, TRT3) for the second experiment. Therefore, the numbers of
samples per group in each experiment are $n_{11}=n_{12}=n_{13}=2$,
$n_{14}=0$ and $n_{21}=n_{22}=n_{23}=1$, $n_{24}=3$. Note that this
experimental design is not as balanced as it might have been, for
example, we could have required at least one sample from each group in
each experiment.  However it might not always be possible to construct
``balanced'' experiments due to having a large number of groups
relative to the number of tags available, and so we investigate here
the effect of using an ``unbalanced'' experimental design.  As we have
a sample from (control) group~1 in each experiment, we take these
samples as the reference samples for the experiments, that is, take
$g_1=g_2=1$.

We constructed the dataset to contain the results on $P=300$ proteins,
with the number of simulated MS/MS spectra per protein ($m_j$) drawn
from a geometric distribution with mean~$6$, this distribution being
chosen to be roughly comparable with the real MS/MS datasets analyzed
in this paper. The mean log-intensities of the reporter ions for
(control) group~1 ($\alpha_{jk}$) were drawn from a normal
distribution with mean $10$ and standard deviation $3.0$.  Again, this
distribution is similar to those for the real datasets in the paper.
The probability of a protein being differentially expressed with
respect to the control group CTL was drawn from Bernoulli
distributions with probabilities $0.1$, $0.2$ and $0.3$ for treatment
groups TRT1, TRT2 and TRT3 respectively and resulted in $26$, $71$ and
$98$ differentially expressed proteins in the treatment groups.  We
considered fold changes in differential expression relative to the
(control) group varying between $1.5$ and $4.0$ for up-regulation and
between $1/4.0$ and $1/1.5$ for down-regulation. This led to us
drawing the parameters for the level of differential expression
($\gamma_{gj}$) from a uniform distribution on $(-1.39,-0.41)\cup
(0.41,1.39)$.  The log ratios of the amount of protein in each sample
with respect to a reference sample for each experiment (sample~1 in
group~$g_e$ of experiment~$e$), the $\kappa_{egi}$, were drawn from a
normal distribution with mean $0$ and standard deviation $0.1$.
Finally the variability of the errors ($\sigma$) was set to~$0.3$.

\begin{figure}[t!]
\includegraphics[angle=270, width=\textwidth]{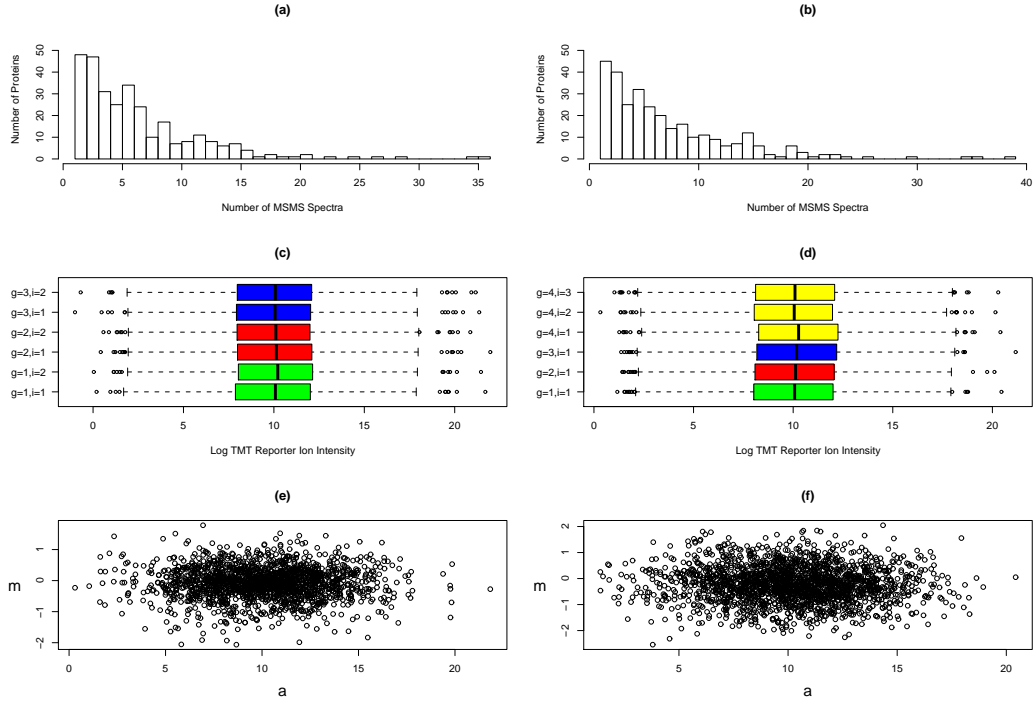}
\caption{Summary plots for the simulated data: left panel --
  experiment~1, right panel -- experiment~2. (a) and (b) Histograms of
  the number of MS/MS spectra per protein; (c) and (d) Box-plots of
  the reporter ion log-intensities by group/sample; (e) and (f)
  MA-plots for two selected samples (showing the difference in
  log-intensities, \textsf{m}, against the average log-intensity,
  \textsf{a}).}
\label{fig:simulated_data_summary}
\end{figure}

The data are summarized in Figure~\ref{fig:simulated_data_summary},
with the left-hand column giving summaries for experiment~1 and the
right-hand column for experiment~2.
Figures~\ref{fig:simulated_data_summary}(a) and~(b) show histograms of
the number of MS/MS spectra per protein for the two experiments.
Clearly the number of proteins decreases as the number of spectra per
protein increases, and is consistent with the simulation model in
which the number of spectra per protein follows a geometric
distribution with mean~6. Figures~\ref{fig:simulated_data_summary}(c)
and~(d) are box plots of the log-intensities of the reporter ions
(labelled by group and sample within group) in the two experiments.
The plots show that there are no clear differences between
groups/samples, with these levels dominated by the overall mean levels
($\alpha_{jk}$), which in this simulation have mean~10 and standard
deviation~3. Figures~\ref{fig:simulated_data_summary}(e) and~(f) are
MA-plots for two selected samples (sample~1 in group~2 (TRT1) of
experiment~1 and sample~1 in group~4 (TRT3) of experiment~2). These
plots display differences between these samples and the reference
sample (in the corresponding experiment) in the log-intensities ($m$)
against their mean ($a$) for each MS/MS spectra in an attempt to
highlight any dependence between variability and overall level in the
data. These plots suggest (correctly) that there is no such dependence
in these data.

\subsection{Results}
\label{sec:results-sim}
We analysed these data using the prior distribution with the default
parameter options and looked at the output of the JAGS code using a
variety of initial starting points. Typically, 10-30K iterations were
needed to attain distributional convergence (burn-in) but for the
analyses in this paper we adopt a conservative strategy and use a
burn-in of 100K iterations. Note that convergence was assessed by
using a variety of informal and formal tests.  We report here the
results from five independent runs of the JAGS code in which (after
burn-in) each chain was run for a further 100K iterations and thinned
by taking every 100th iterate. This gives a total of 5K (weakly
correlated) realizations from the posterior distribution for analysis.

\begin{figure}[h]
\begin{center}
\includegraphics[angle=270, width=5.5in]{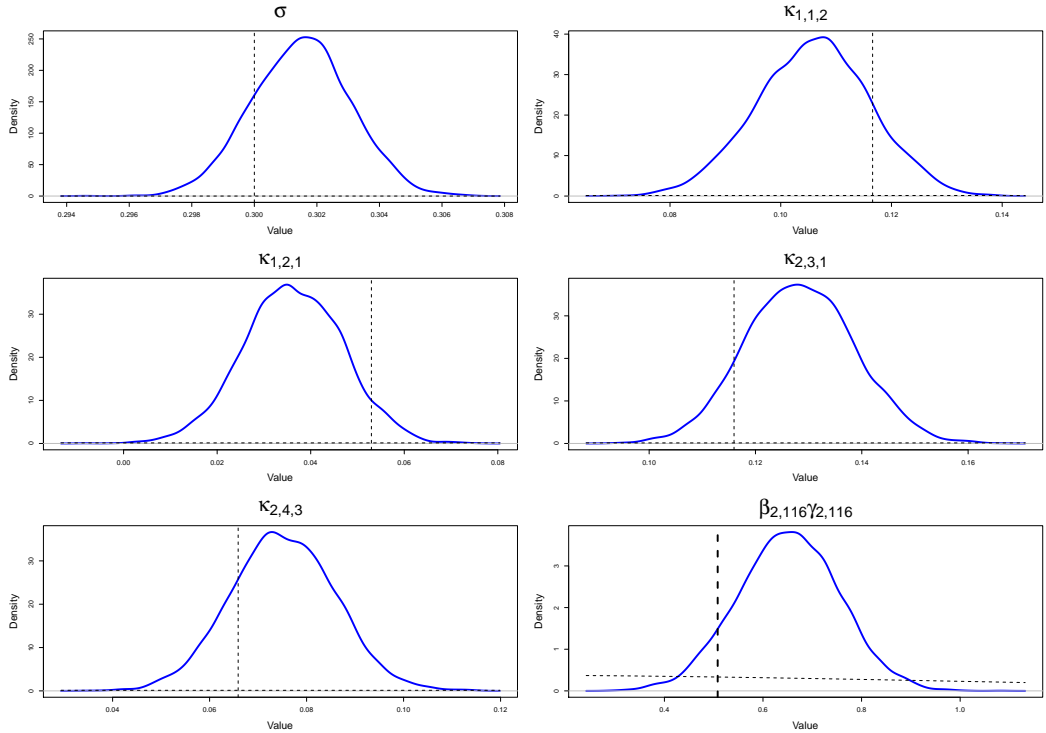}
\caption{Posterior kernel density plots for a selection of parameters:
  the measurement error standard deviation $\sigma$, four of the
  normalisation constants $\kappa_{egi}$ and a log-fold difference in
  intensity with respect to (control) group~1 $\beta_{gj} \gamma_{gj}$
  for a protein which was differentially expressed in this group.  The
  vertical dashed lines indicate the value used in generating the
  simulated data and the dashed curves are the prior densities used in
  the analysis.}
\label{fig:simulated_all_mcmc}
\end{center}
\end{figure}

Figure \ref{fig:simulated_all_mcmc} shows kernel density plots for an
MCMC sample from the posterior probability distribution for a
representative selection of parameters: the measurement error standard
deviation $\sigma$, four of the normalisation constants $\kappa_{egi}$
and a log-fold difference in intensity with respect to (control)
group~1 $\beta_{gj} \gamma_{gj}$ for a protein which was
differentially expressed in this group.  The vertical dashed lines
show the value used in generating the simulated data. Their location
on the posterior plots verifies that, despite inputting fairly weak
prior information, the posterior analysis has recovered these values
reasonably accurately.  Note that there is no ``spike'' at zero in the
posterior density of $\beta_{2,116}\gamma_{2,116}$ because there were
no zero values of $\beta_{2,116}$ in the posterior sample.

\begin{figure}[t!] 
\centerline{
\includegraphics[angle=270,width=0.33\textwidth]{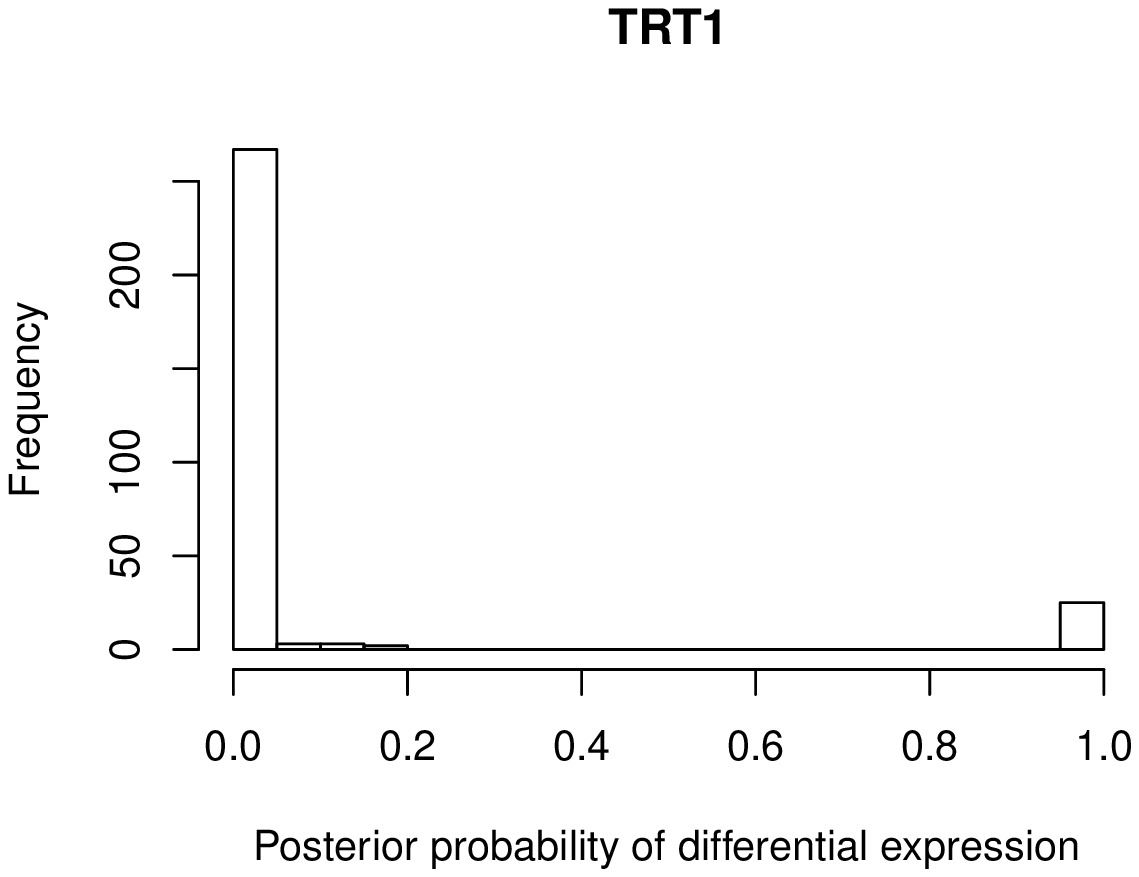}
\includegraphics[angle=270,width=0.33\textwidth]{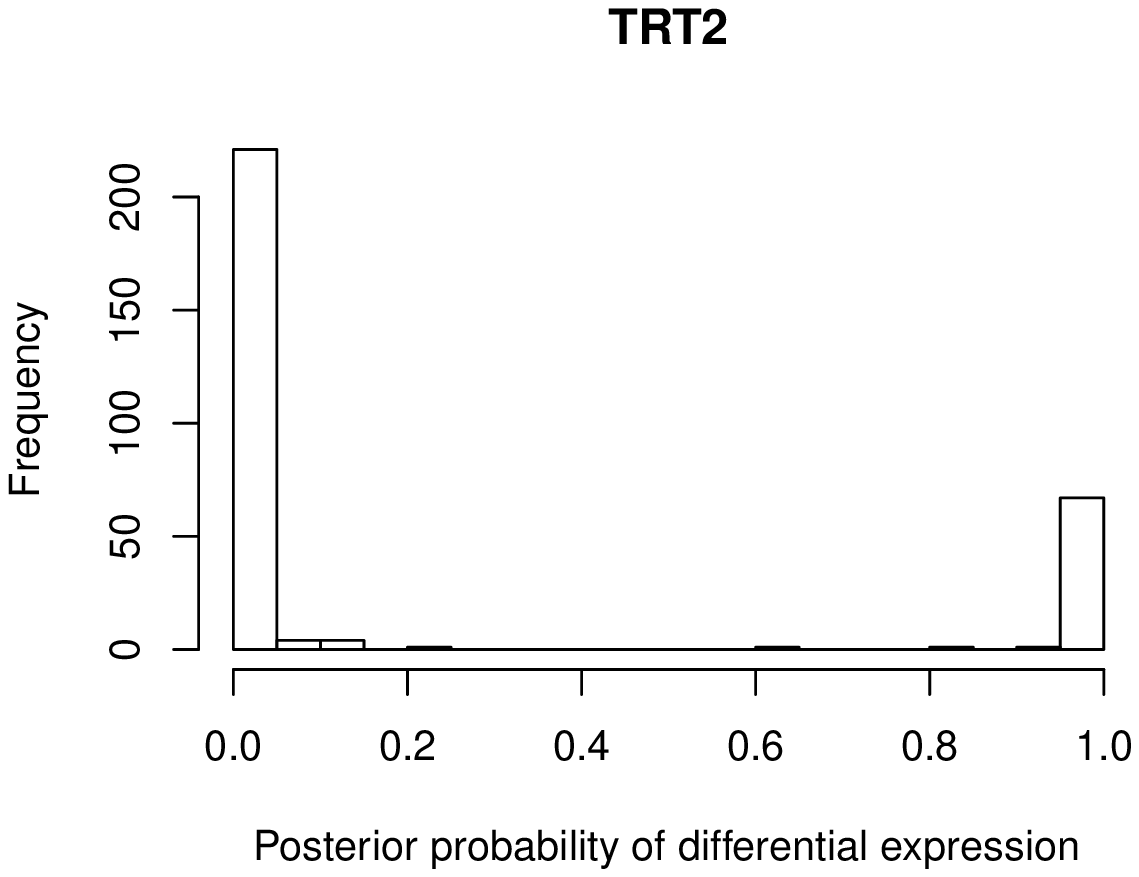}
\includegraphics[angle=270,width=0.33\textwidth]{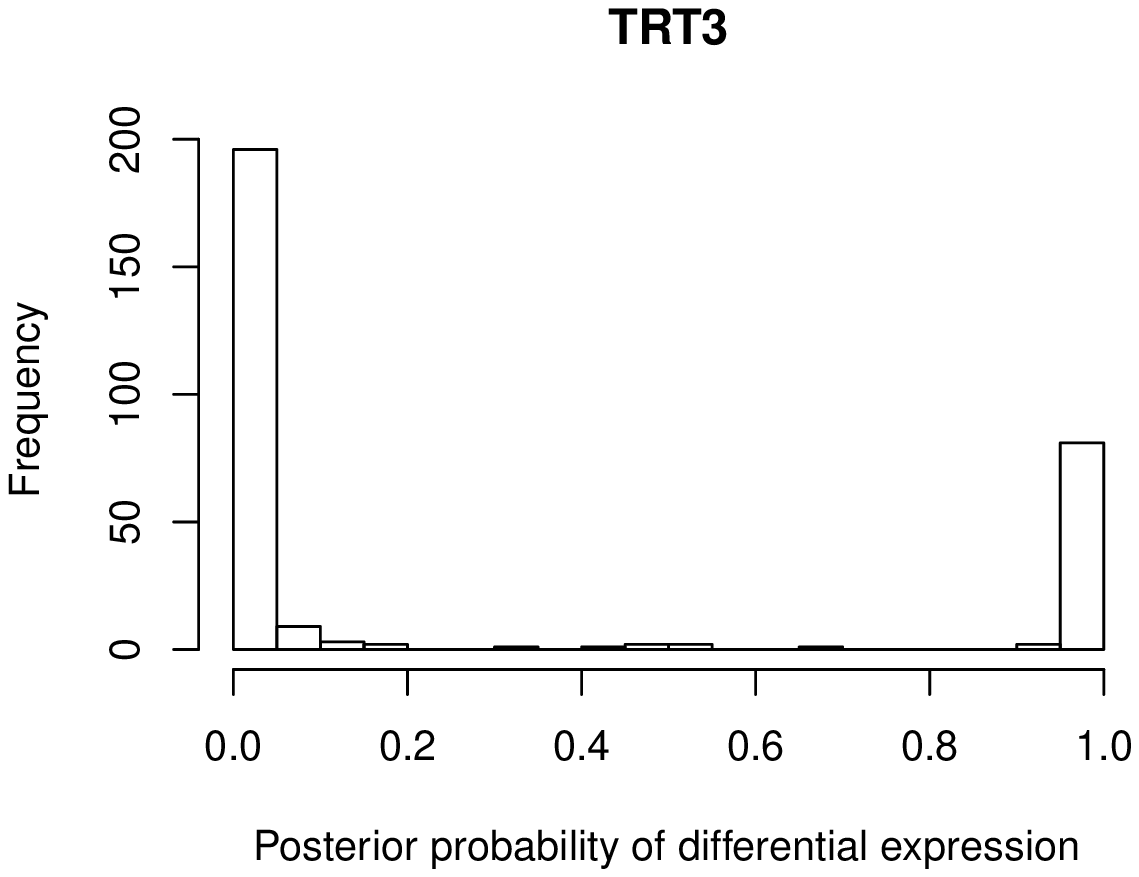}
}
\caption{Histograms showing the posterior probability of differential expression for the proteins in groups TRT1, TRT2 and TRT3, respectively. The vast majority of proteins in the 3 samples have posterior probability very close to 0 or 1.}
\label{fig:sim:hist}
\end{figure}

Table~\ref{tab:simulated_TRT_contingency} summarises the overall
performance of the inference method in identifying the differentially
expressed proteins. Classification is based on a posterior probability
of differential expression exceeding 0.5, as is common practice for
Bayesian variable selection problems. In fact the classification is
rather insensitive to the particular choice of probability threshold,
since the vast majority of proteins have posterior probability of
differential expression either less than 0.1 or greater than 0.9. The
full distribution of posterior probability of differential expression
for the proteins in each of the 3 treatment groups is summarised in
Figure~\ref{fig:sim:hist}. Despite the relatively high noise level,
the method performs quite well for TRT1 and TRT2 but less well for
TRT3.  On further inspection, we found that the method generally
failed to correctly identify a differentially expressed protein when
either the level of differential expression was small or there were
only one or two MS/MS spectra for the protein in each experiment.
Additional complications for TRT3 were due to its samples only being
present in the second experiment.

\begin{table}[t!]
\begin{center}
\begin{tabular}{|l|c|c|c|}
\cline{2-3}
\multicolumn{1}{l|}{Group TRT1} &  Simulated   & Simulated  \\
\multicolumn{1}{l|}{~}  &  with DE     & without DE \\
\hline 
Inferred as with DE     &     24       & \d\d 1     &\d 25 \\ 
Inferred as without DE  &  \d  2       &    273     &  275 \\
\hline 
\multicolumn{1}{l|}{~}  &     26       &    274     &  300 \\
\cline{2-4} 
\noalign{\bigskip}
\cline{2-3}
\multicolumn{1}{l|}{Group TRT2} &  Simulated   & Simulated  \\
\multicolumn{1}{l|}{~}  &  with DE     & without DE \\
\hline 
Inferred as with DE     &     70       & \d\d 0     &\d 70 \\ 
Inferred as without DE  &  \d  1       &    229     &  230 \\
\hline 
\multicolumn{1}{l|}{~}  &     71       &    229     &  300 \\
\cline{2-4} 
\noalign{\bigskip}
\cline{2-3}
\multicolumn{1}{l|}{Group TRT3} &  Simulated   & Simulated  \\
\multicolumn{1}{l|}{~}  &  with DE     & without DE \\
\hline 
Inferred as with DE     &     87       & \d\d 0     &\d 87 \\ 
Inferred as without DE  &     11       &    202     &  213 \\
\hline 
\multicolumn{1}{l|}{~}  &     98       &    202     &  300 \\
\cline{2-4} 
\end{tabular}
\caption{Performance of the method in detecting differential
  expression (DE) in proteins between groups and the control group.
  Inference for DE is based on a classification threshold of a
  posterior probability exceeding 0.5. }
\label{tab:simulated_TRT_contingency}
\end{center}
\end{table}

\section{Case Study 1: Dataset on human plasma}
\label{sec:plasma}
Our first case study analyses data from an MS/MS analysis of human
plasma published in \citet{Dayon2010}.  The data arise from an
experiment to investigate the use of a particular mass spectrometric
technique in the identification and quantification of peptides. The
experimental design consists of a single experiment ($E=1$) and
produced two technically identical samples.  We consider these samples
to be single samples ($n_{11} = n_{12} = 1$) from two ``artificial''
groups ($G=2$). Thus, any differentially expressed proteins will be
due to variability in the experimental process and the technique used
to detect differential expression, and so we expect to find relatively
few proteins that are differentially expressed between the two groups.

The experiment was conducted as follows. After reduction, alkylation
and digestion of the sample with trypsin, two technically identical
sub-samples were taken, labelled with TMT-2plex labels (i.e.  $n_I=2$)
and mixed. The resulting mixture was then run through a liquid
chromatograph into a Tandem Mass Spectrometer and MS/MS spectra
acquired.  In this simple experiment, it does not really matter which
group supplies the reference sample. Here, we take this as coming from
group~1, that is, $g_1=1$. The MS/MS spectra, in the form of Mascot
Generic Format (MGF) files, were kindly made available for this study
by Alexander Scherl.  These were then analysed using Proteome
Discoverer version $1.1$ (Thermo Fisher Scientific) and version $3.65$
of the IPI sequence database \citep{IPI2004}. The list of identified
MS/MS spectra was then filtered by requiring them to pass three
criteria: (i) the peptide needed a high identification confidence,
with the threshold calculated to give a false discovery rate on the
decoy database of $0.01$, (ii) the MS/MS spectra needed to be
identified as being those of peptides uniquely derivable from a single
protein, and (iii) the MS/MS spectra needed to include both reporter
ion peaks.  This left $3158$ MS/MS spectra, corresponding to peptides
from $P=94$ proteins, for analysis.

\begin{figure}[t!]
\includegraphics[angle=270,width=\textwidth]{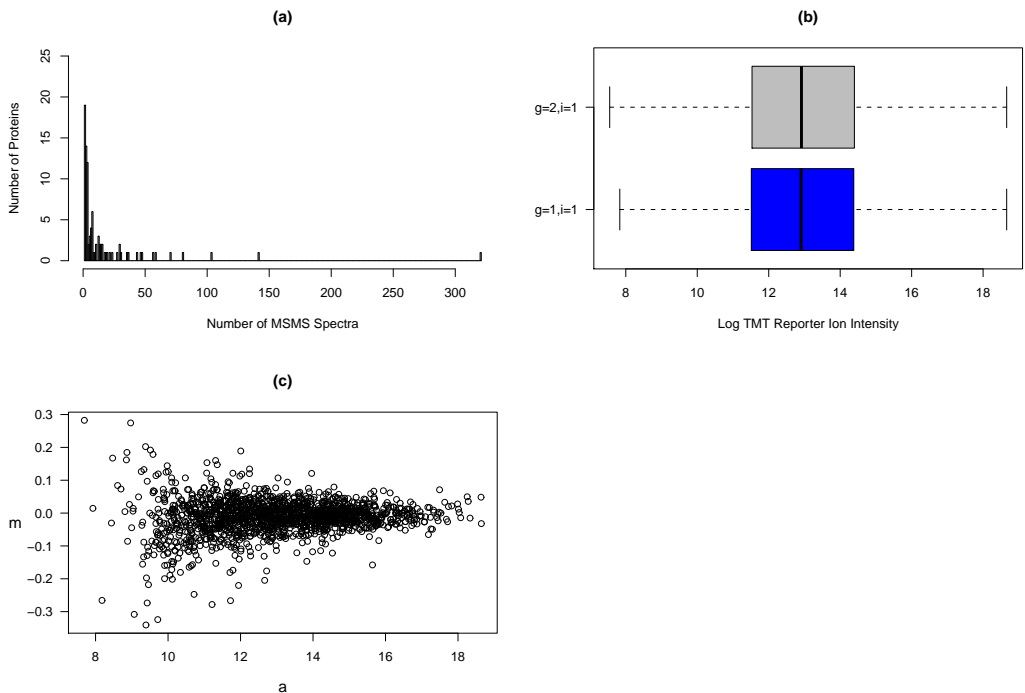}
\caption{Summary plots of the \citet{Dayon2010} human plasma data. (a)
  Histogram of the number of MS/MS spectra per protein; (b) Box-plot
  of the TMT 2-Plex reporter ion log-intensities by group/sample; (c)
  MA-plot of the log-intensities for the two samples (showing the
  difference in log-intensities, \textsf{m}, against the average
  log-intensity, \textsf{a}).}
\label{fig:dayon2010_peptide_distribution}
\end{figure}

Figure~\ref{fig:dayon2010_peptide_distribution} gives some summary
plots of the data. Figure~\ref{fig:dayon2010_peptide_distribution}(a)
gives a histogram of the number of MS/MS spectra per protein. Clearly
there are many proteins with only a few spectra and only a few with
many spectra. Note that this shape is consistent with the geometric
distribution assumed in the simulation study.
Figure~\ref{fig:dayon2010_peptide_distribution}(b) gives a box plot of
the log-intensities of the reporter ions (labelled by group and sample
within group) and shows no obvious differences between the groups.
Figure~\ref{fig:dayon2010_peptide_distribution}(c) gives the MA-plot
for the two samples. The plot is similar to those in the simulation
study (Figures~\ref{fig:simulated_data_summary}(e) and~(f)). Here
there is a slight suggestion of increased variability associated with
small measurements, but the effect seems to be small, and so for the
purposes of this analysis we will continue to assume constant variability.

\subsection{Results}
\label{sec:plasma_results}
We now analyse these duplex TMT reporter ion log-intensities and
report here the results from runs of the JAGS code using the procedure
described in section~\ref{sec:results-sim} which give 5K (weakly
correlated) realizations from the posterior distribution.

\begin{figure}[th!]
\begin{center}
\includegraphics[angle=270,width=\textwidth]{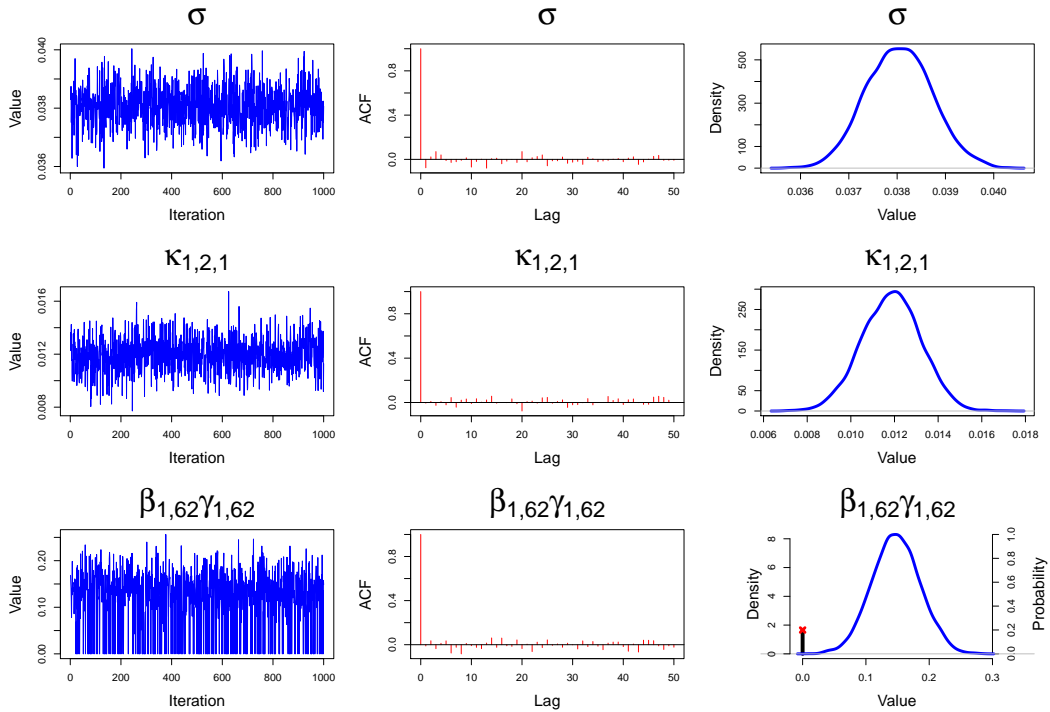}
\caption{Results for an analysis of the data in \citet{Dayon2010}:
  trace and auto-correlation plots from a single chain and the
  posterior kernel density plot from all five chains for a
  representative selection of parameters: the measurement error
  standard deviation $\sigma$, the log-normalisation parameter
  $\kappa_{1,2,1}$ and the product $\beta_{2,62}\gamma_{2,62}$ for
  protein IPI00647704.}
\label{fig:dayonmcmc}
\end{center}
\end{figure}

Figure \ref{fig:dayonmcmc} shows summary plots for an MCMC sample from
the posterior probability distribution for a representative selection
of parameters: the measurement error standard deviation $\sigma$, the
(only) normalisation constant $\kappa_{1,2,1}$ for the second sample
($g=2,i=1$) with respect the reference sample ($g=1,i=1$) and the
log-fold difference in intensity $\beta_{2,62}\gamma_{2,62}$ for
protein~IPI00647704 with respect to (control) group~1. This protein
was chosen as it has the highest posterior probability of being
differentially expressed.  The trace and auto-correlation plots are
typical of all five chains and suggest that there are no mixing
problems and that convergence has been attained.

Means of the posterior output for the differential expression
indicator parameters $\beta_{gj}$ can be used to estimate the
posterior probability of differential expression for each protein.  Of
the $94$ proteins examined, only two had a posterior probability
exceeding $0.1$. These were IPI00647704 and IPI00916434, with
posterior probabilities $0.8$ and $0.3$ respectively. If we assume
that the two samples were perfect technical replicates, then using a
posterior probability threshold of $0.5$ to declare whether or not
proteins are differentially expressed gives a false positive rate
around~$1\%$. 

\begin{figure}[ph!]
\begin{center}
\includegraphics[scale=0.7]{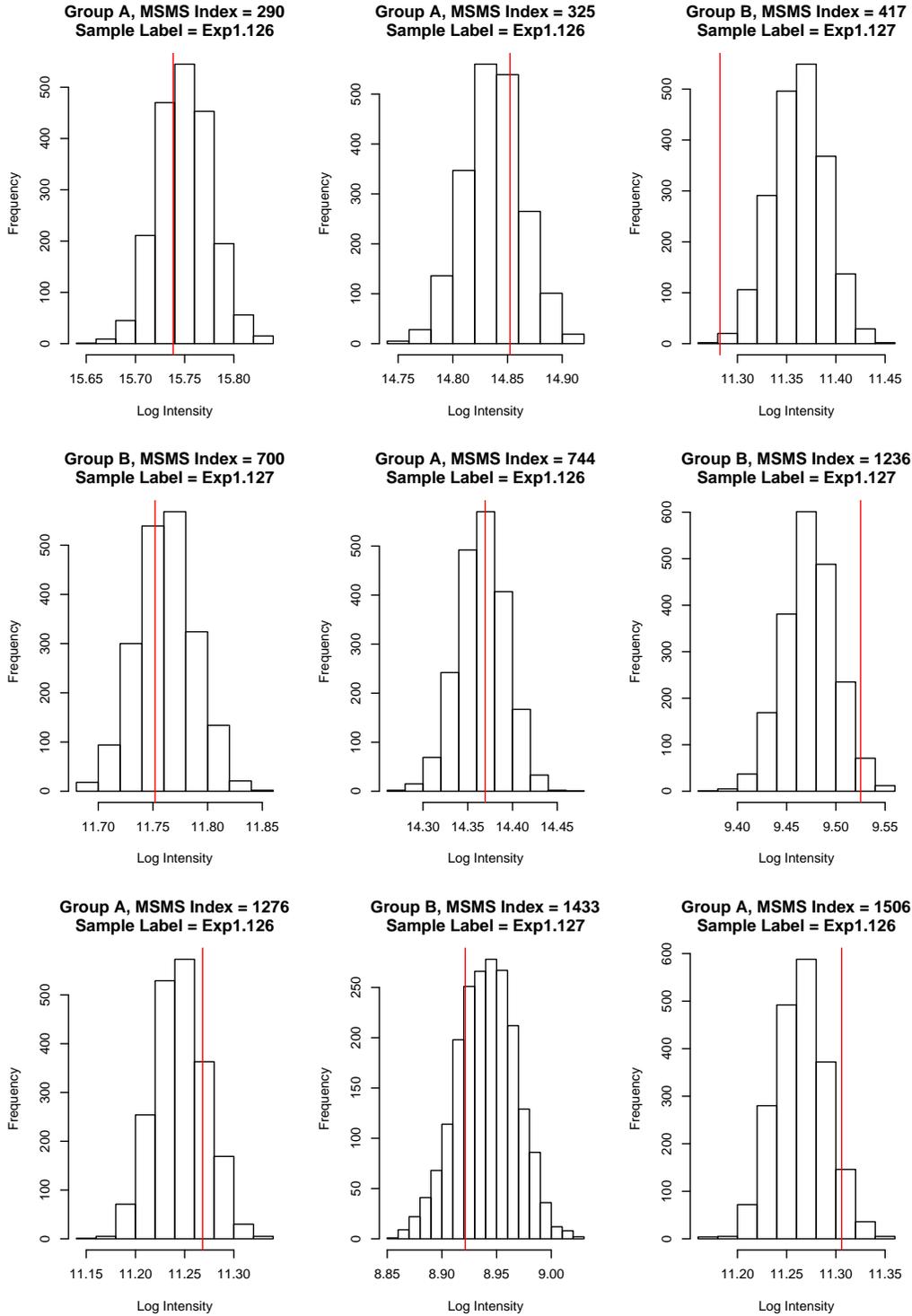}
\caption{Histograms showing the posterior predictive distributions for
  a random selection of observed log-intensities (uniquely
  identifiable by experiment $e$, protein~$j$, MSMS spectra $k$ and
  sample $gi$). The red vertical line shows the observed
  log-intensity.}
\label{fig:dayon_pit}
\end{center}
\end{figure}

A natural part of any data analysis is to assess the validity
  of the model used to make inferences. We favour checks using the
  posterior predictive distribution of the (logged) intensities, that
  is, their distribution allowing for the posterior uncertainty in the
  model parameters; see \cite{Gelman03}, Chapter~6 for details.  The
  posterior predictive density is straightforward to determine using
  the MCMC output
  $\{\kappa_{egi}^{(\ell)},\alpha_{jk}^{(\ell)},\beta_{gj}^{(\ell)},\gamma_{gj}^{(\ell)},\sigma^{(\ell)};~\ell=1,\ldots,N\}$
  for our model as
\[
f\left(y_{egjki}\right)
\simeq\frac{1}{N}\sum_{\ell=1}^N \phi\left(\frac{y_{egjki}-(\kappa_{egi}^{(\ell)}+\alpha_{jk}^{(\ell)}+\beta_{gj}^{(\ell)}\gamma_{gj}^{(\ell)})}{\sigma^{(\ell)}}\right).
\]
A useful diagnostic of model fit can be based on the location of the
observed intensities within their individual predictive distributions
as a well fitting model will produce posterior predictive
distributions consistent with the observed log-intensities.
Figure~\ref{fig:dayon_pit} shows histograms of samples from the
posterior predictive densities for a random selection of nine observed
log-intensities.  These show that the model provides a good fit to the
data (shown in red).

It is interesting to compare our results with those obtained by using
existing methods: a t-test and a moderated t-test \citep{Smyth2004},
both using log intensity data, as in our Bayesian analysis. Since the
t-test looks at differences between peptide intensities (on the log
scale), the normalisation constant for specific peptides
(corresponding to $\alpha$ in the Bayesian model) drops out of the
analysis, and therefore does not require explicit normalisation.
However, the normalisation constant associated with the sample
(corresponding to $\kappa$ in the Bayesian model) does not drop out,
and therefore the raw log intensities must be pre-processed in order
to apply sample normalisation before the t-test can be carried out.
Here we normalise each sample by its mean value.  This is one of the
standard normalisation techniques commonly used, and is the method
most directly comparable to our Bayesian model. Including the
normalisation constants explicitly in the model has the advantage that
it allows direct modelling of the experimental data. Our approach to
modelling normalisation constants is similar to the approach used by
\cite{Oberg2008} in a frequentist context. For these data, both the
t-test and the moderated t-test suggest that there are no
differentially expressed proteins at the $5$\% significance level
after using an FDR correction for testing multiple hypothesis
\citep{Benjamini1995}. It is worth noting that the pre-normalisation
step is necessary, and that different methods of normalisation can
lead to different results \citep{karpievitch2012normalization}.  By
contrast, the Bayesian model-based method developed here includes
normalization constants explicitly as part of an integrated model and
so normalizing the data prior to analysis is not required. We have
also found that our method is relatively insensitive to whether or not
the data have been preprocessed using standard normalization methods.

\section{Case Study 2: ProteoRed multi-centric experiment~5}
\label{sec:proteoRed}
Our second case study analyses data from an experiment carried out by
the ProteoRed consortium as part of an assessment of various
quantitative proteomics methods. The experiment analysed a mixture of
\emph{E.\ coli} proteins.  The mixture was prepared by fractionation
of the cytoplasmic proteome of \emph{E.\ coli} and contained soluble
proteins with a wide range of values for the isoelectric point (pI,
the pH at which a molecule carries no net electrical charge) and
average molar mass (Mw).

After dividing the mixture into two identical portions A and B, four
xenobiotic proteins were added into the portions in differing amounts.
The aim of this research was to evaluate the performance of different
proteomics methods by observing their ability to detect the existence
of the xenobiotic proteins within the otherwise identical
samples/portions. The xenobiotic proteins used were CYC\_HORSE
(Cytochrome C, Mw 12362), MYG\_HORSE (Apomyoglobin, Mw 16952),
ALDOA\_RABIT (Aldolase, Mw 39212) and ALBU\_BOVIN (Serum albumin, Mw
66430) and their (theoretical) differential expression between the two
portions is given in Table~\ref{tab:proteored_xenobiotic}.

\begin{table}[h]
\begin{center} 
\begin{tabular}{|l|l|c|c|}\hline
 Code & Protein          & Ratio ($B/A$) & $\log(B/A)$ \\\hline 
CYC   &  Horse Cytochrome C   & $1.50$ &  \phantom{$-$}$0.4055$\\
MYG   &  Horse Apomyoglobin   & $0.38$ &  $-0.9676$\\
ALDOA & Rabbit Aldolase       & $0.50$ &  $-0.6931$\\
ALBU  & Bovine Serum Albumin  & $5.00$ &  \phantom{$-$}$1.6094$ \\\hline
\end{tabular}
\caption{Theoretical ratios of xenobiotic proteins in portions A and B}
\label{tab:proteored_xenobiotic}
\end{center}
\end{table}

\begin{figure}[ht!]
\includegraphics[angle=270, width=\textwidth]{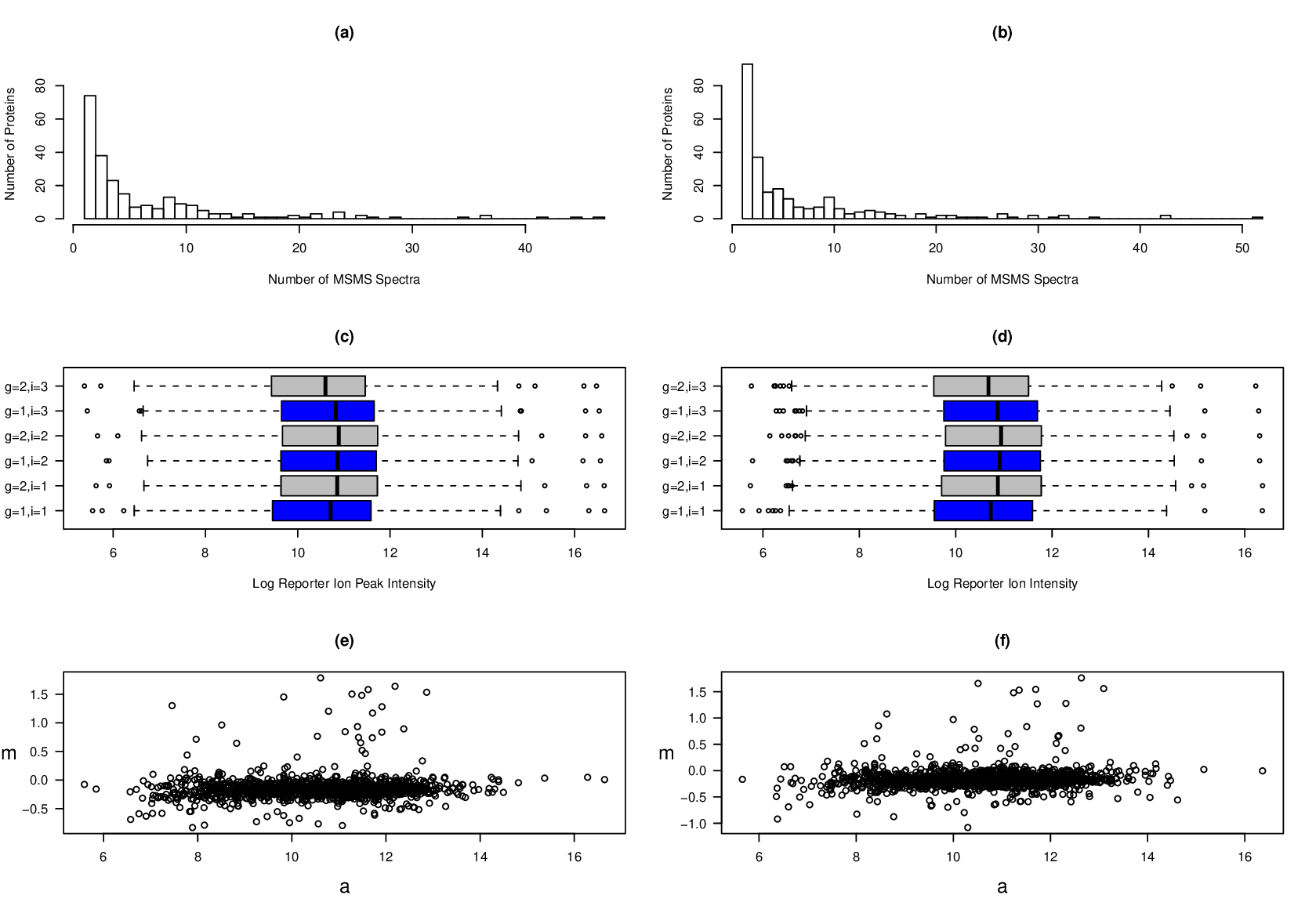}
\caption{Summary plots for the ProteoRed data: left panel --
  experiment~1, right panel -- experiment~2. (a) and (b) Histograms of
  the number of MS/MS spectra per protein; (c) and (d) Box-plots of
  the reporter ion log-intensities by group/sample; (e) and (f)
  MA-plots for two selected samples (showing the difference in
  log-intensities, \textsf{m}, against the average log-intensity,
  \textsf{a}).}

\label{fig:proteored_all_peptide_distribution}
\end{figure}

\begin{figure}[ht!]
\includegraphics[angle=270,width=\textwidth ]{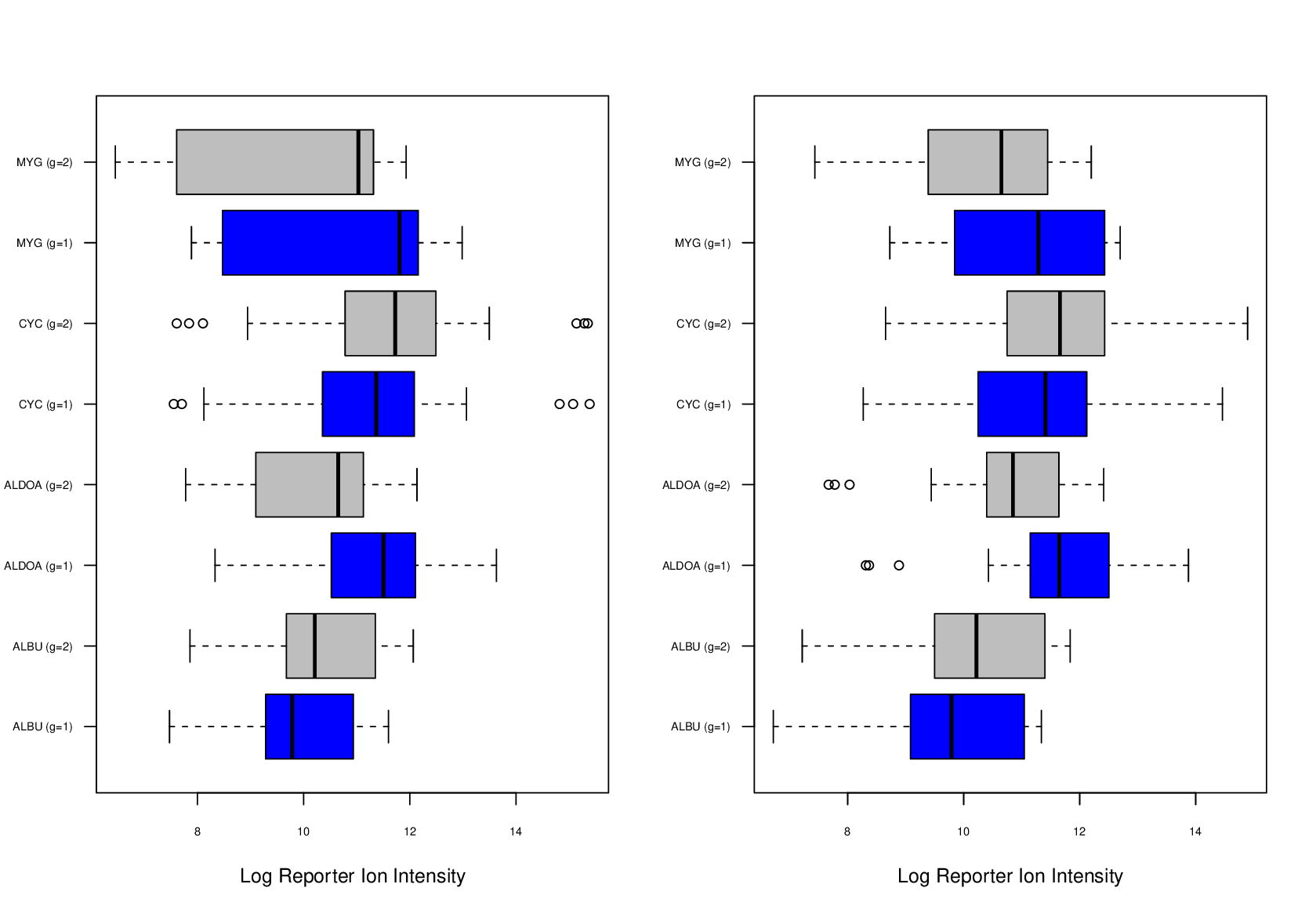}
\caption{Box-plots of the reporter ion log-intensities by group for
  the four spiked in proteins: left panel -- experiment~1, right panel
  -- experiment~2.}
\label{fig:proteored_spiked_peptide_distribution}
\end{figure}

The experiment produced data by using a TMT-6plex ($n_I=6$) labelling
of three technical replicate sub-samples of each of the two portions.
Sub-samples in portion A were labelled with even TMT-6plex labels
(TMT126, TMT128, TMT130) and, in portion B, labelled with odd
TMT-6plex labels (TMT127, TMT129, TMT131). The resulting labelled
sub-samples were then mixed and the mixture divided into two
technically identical portions. Each of these portions was then
subjected to independent MS/MS analyses. In terms of our model, this
gives an experimental design consisting of $E=2$ experiments and $G=2$
groups, and we label portion~A as group~1 and portion~B as group~2.
The numbers of replicates in each experiment$\times$group are
$n_{11}=n_{12}=n_{21}=n_{22}=3$.  The reference sample was set to its
default in both experiments ($g_1=g_2=1$).

We obtained the MS/MS peak list as an MGF file for both of the
analyses. These were then analysed using Proteome Discoverer v.1.1
(Thermo Fisher Scientific) and the current version of the UniProt
sequence database for \emph{E.\ coli} \citep{Uniprot2009}. The list of
identified peptides was then filtered to retain only those peptides
with a high identification confidence, with the threshold calculated
to give a false discovery rate on the decoy database of $0.01$, and
which were proteotypic. Also MS/MS spectra which did not have complete
quantitative information for the six labels were excluded from the
analysis.  This left $238$ and $259$ proteins, each with at least one
fully quantified MS/MS spectra, for the analyses of portions~A and~B
respectively. In total, the two analyses provided quantitative
information for $282$~proteins. The data are summarised in
Figures~\ref{fig:proteored_all_peptide_distribution}
and~\ref{fig:proteored_spiked_peptide_distribution}.
Figures~\ref{fig:proteored_all_peptide_distribution}(a) and~(b) shows
histograms of the number of MS/MS spectra per protein for each
experiment. As with the first case study, these distributions are
consistent a geometric distribution.
Figures~\ref{fig:proteored_all_peptide_distribution}(c) and~(d) are
box plots of the log-intensities of the reporter ions (labelled by
group and sample within group), and panels~(e) and~(f) show MA plots for two
of the samples. All of the plots suggest that the data is consistent
with the assumptions of our model.

\subsection{Results}
We now analyse the ProteoRed dataset and, as with the previous
analyses, base our analysis on the 5K realizations from the posterior
distribution obtained by using the procedure described in
section~\ref{sec:results-sim}.

\begin{figure}[t]
\includegraphics[angle=270, width=\textwidth]{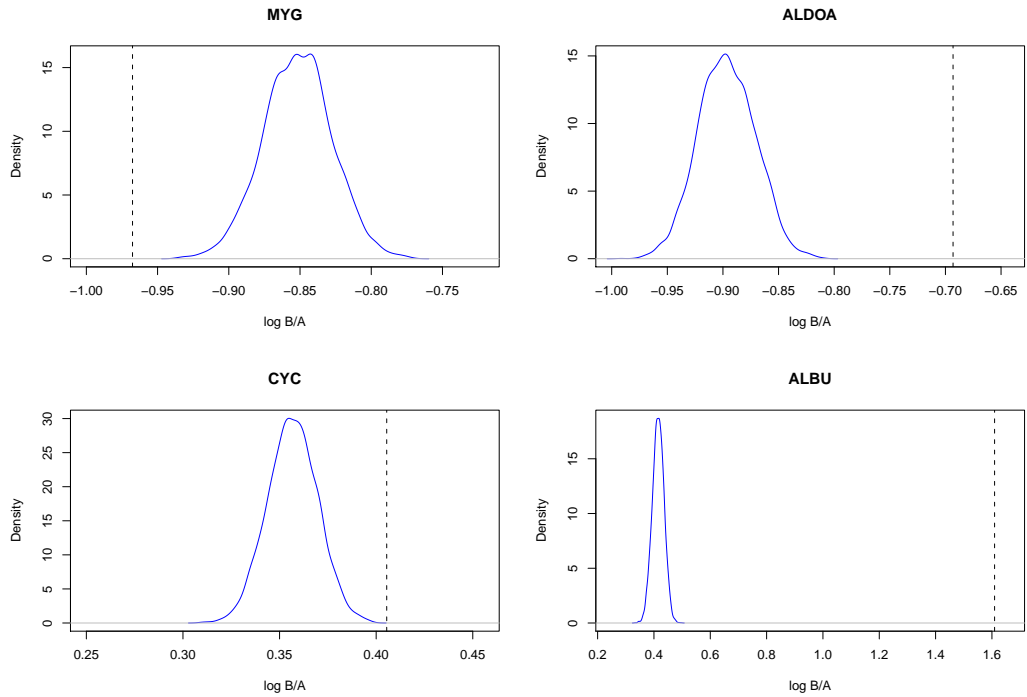}
\caption{Posterior kernel density plots for the log ratio $\log(B/A)$
  of the spiked proteins MYG, ALDOA, CYC and ALBU. The vertical dashed
  line shows the theoretical value for the log ratio.}
\label{fig:proteored_spiked_protein_ratios}
\end{figure}

The analysis clearly identified all four spiked-in proteins as being
differentially expressed, each with a posterior probability
exceeding~$0.999$. The posterior distribution of the log ratio
$\log(B/A)$ for each spiked protein is shown in
Figure~\ref{fig:proteored_spiked_protein_ratios}. Comparing these
distributions with the theoretical values given in
Table~\ref{tab:proteored_xenobiotic}, we see that the values of the
distributions for three of the four spiked proteins are of the right
order of magnitude. Unfortunately, the analysis gives a posterior
ratio for the ALBU (Bovine Serum Albumin) spiked protein which is out
by a factor of more than~3. Other analyses of this dataset have come
to similar conclusions (see results section in \citet{ProteoRed5}) and
we suspect this is due to contamination in the original sample or at
some stage in the preparation of the samples for the MS/MS experiment.
Whatever the reason, this has clearly affected the estimation of the
ratios for all the spiked proteins.

The analysis also identified four other (non-spiked) \emph{E.\ coli}
proteins as being differentially expressed when using a posterior
probability threshold of~$0.5$: three main ones A7J0Q2, C3SFD7 and
P02754 with posterior probabilities exceeding~$0.999$ and C3SVU2 with
posterior probability~0.52. Nearly all of the remaining \emph{E.\
  coli} proteins ($271$ of the remaining~$274$) had a posterior
probability of being differentially expressed below~0.05. As in the
analysis of the dataset in section~\ref{sec:plasma}, we assessed model
fit by comparing the observed data with their posterior predictive
distributions and this confirmed that the model provided a reasonable
fit to the data.

As before, we now compare the results of our analysis with those
determined by using a standard t-test and by using a moderated t-test
together with an FDR multiple hypotheses correction. Again, we follow
standard practice and pre-normalise the data so that the sample means
are the same.  For both test-based analyses, the spiked proteins were
found to be significantly differentially expressed ($q\leq 10^{-6}$).
These analyses also found false positives, with the t-test analysis
and the moderated t-test analysis identifying an additional ten and
seven non-spiked proteins respectively ($q\leq 0.05$).

\section{Discussion}

In this paper we developed a model based approach to solving a general
problem in quantitative proteomics using mass spectrometric (MS/MS)
methods with isobaric labelling.  The problem is that of determining
which, if any, of a set of proteins in a complex mixture are
differentially expressed between two or more groups and if so to
quantify the degree of differential expression. This is not
straightforward to resolve given the inherently stochastic nature of
mass spectrometric data. Further complications are introduced, given
that the protein intensities are not measured directly. What is
observed are the intensities of the isobaric labels for the peptides
resulting from enzyme digestion (usually trypsin) of these proteins.

The model based approach allows us to integrate data from multiple
MS/MS experiments into a single statistical inference framework.  An
important feature of our framework is that these experiments do not
have to be simple replicate experiments.  This allows us to design a
set of isobaric labellings that expands the effective number of
samples (and therefore biological groups) that can be analysed beyond
the maximum of eight isobaric labels which is currently commercially
available \citep{Choe2007}.  This ability to integrate data is a
distinct advantage not offered by more standard approaches. Previous
work by \citet{Hill2008} also uses a model based approach and makes
this very point about combining data from multiple experiments.
However, we use a more flexible and, we believe, more intuitively
interpretable Bayesian statistical approach to inferring model
parameters, whereas they use an ANOVA-based approach.  An additional
advantage of the Bayesian approach is that we can make direct
inferences for the probability of differential expression, and so
there is no need to correct for multiple hypotheses, as would be
required for the ANOVA approach, for example, in order to determine
whether protein ratios differed significantly from one.

A common concern with Bayesian analyses can be that results are
sensitive to the choice of prior distribution parameters. In order to
assess whether our results were robust and not an artifact of the
prior choice, we investigated using a wide range of different values
of the parameters of the priors.  We found that the results for
differentially expressed proteins were very robust for different prior
choices for $\tau$ and $\kappa$. The results are slightly more
sensitive to the prior choices for the distribution on $p_{gj}$, the
prior probability of protein $j$ in group $g$ being differentially
expressed. Sensitivity to the prior on variable inclusion is a
well-known issue in Bayesian variable selection \citep{Hara2009}.

In order to evaluate the method developed here we apply the technique
to two MS/MS datasets.  First the performance of the method was
evaluated on a negative control. This is a dataset which we know
should have no differentially expressed proteins. The dataset was from
an MS/MS analysis of two technical replicate samples of human plasma
labelled with a 2-plex TMT reagent. For this dataset we can reasonably
assume that none of the proteins are differentially expressed. Of the
$94$ proteins identified from the data only a single protein was
inferred as having a $\geq 0.5$ mean posterior probability of being
differentially expressed with a value of $0.815$. This corresponds to
an acceptably low false positive rate.

The second dataset was data from an MS/MS analysis of a sample used in
a ProteoRed group experiment. The data consisted of technical
replicates of an \emph{E.\ coli} cytoplasmic proteome sample (A and
B). These were spiked with differing amounts of four xenobiotic
proteins. This acted as our positive control since the spiked proteins
and their relative amounts are known. Again the method developed here
was able to infer that the spiked proteins were differentially
expressed with a high probability.  The degree differential expression
for these proteins was also of the right order for three of the
proteins but was off by a factor $>3.0$ for the Bovine Serum Albumin
(ALBU) protein.  Four out of the $278$ non-spiked proteins were also
inferred as differentially expressed: three with a high probability
($>0.999$) and one with a borderline probability of $0.52$.

In conclusion, our Bayesian statistical inference approach to
determining differentially expressed proteins from isobaric labelled
MS/MS data has been found to perform well on a variety of real and
simulated datasets.  The modelling framework allows us to perform a
unified statistical analysis of multiple experiments and the multiple
hypotheses are integrated into the model itself so there is no need
for \emph{ad hoc} normalisation methods or multiple hypothesis
corrections. The model allows comparison of multiple experiments
within a single unified model, thereby extending the range of
applicability of the isobaric labelling technology.  Importantly the
model has a variable selection form which ensures that we fit an
appropriate model for the combination of differentially expressed and
non-differentially expressed proteins.  Models without this structure
have the drawback that they inflate the error variance in the ``null''
model due to contamination by outlying differentially expressed
proteins which then hinders the detection of differential expression.
A unified analysis approach using a frequentist ANOVA analysis has
previously been described by \citet{Oberg2008}. Fortunately, missing
data is less of an issue when analysing isobaric labelled samples in
MS/MS experiments than it is for other proteomic technologies, such as
LC--MS. Nevertheless, missing values are often present in data sets,
and this can complicate frequentist analysis considerably.  However,
adopting a Bayesian approach is helpful in that it allows us to
marginalise the model over any missing data as a routine part of the
analysis. It would be relatively straightforward to extend our model
to cover informative missingness, but we believe that our current
model captures the most important sources of variation in typical
labelled MS/MS datasets. Finally, the results we obtain are
intuitively interpretable through simple probabilities of the analysed
proteins being differentially expressed.

\subsection*{Acknowledgements}

This work was funded by the UK Biotechnology and Biological Sciences
Research Council through grants BBF0235451 and BBC0082001. The authors
would like to thank Alexander Scherl (Proteomics Core Facility and
Biomedical Proteomics Research Group, University of Geneva) for
providing much of the data analysed in this paper, and to Satomi Miwa,
Achim Treumann and Thomas von Zglinicki (Institute for Ageing and
Health, Newcastle University) for helpful discussions. We would also
like to thank the reviewers who have helped to improve the clarity of
the paper.

\bibliographystyle{bepress}
\bibliography{Isobaric}    

\section*{Appendix}

The JAGS model implementing the differential protein expression model
described in section \ref{sec:model} is shown below.

{\small
\begin{verbatim}
model {     
 Tau ~ dgamma(a.tau,b.tau)   
 Sigma <- 1/sqrt(Tau) 

 for (j in 1:P) {
  for (k in 1:m[j]) {
   Alpha[moffset[j]+k] ~ dnorm(a.alpha,b.alpha)
  }
 }
  
 for (e in 1:E) {
  kappa[e,1] <- 0
  for (s in 2:numberOfSamples[e]) {
   kappa[e,s] ~ dnorm(a.kappa,b.kappa)
  }
 }
  
 for (j in 1:P) {   
  p[1,j] <- 0.0  
  Beta[1,j] <- 0.0
  Gamma[1,j] <- 0.0
  for (g in 2:G) {
   p[g,j] ~ dbeta(a.p,b.p)      
   Beta[g,j] ~ dbern(p[g,j]) 
   Gamma[g,j] ~ dnorm(a.gamma,b.gamma)
  }  
      
  for (n in 1:N[j]) {                 
   intensity[offset[j]+n] ~ dnorm(kappa[experiment[offset[j]+n],
     sample[offset[j]+n] - sampleoffset[experiment[offset[j]+n]]]
      + Alpha[peptide[offset[j]+n]]
      + Beta[group[offset[j]+n],j]*Gamma[group[offset[j]+n],j]
     ,Tau);
  }           
 }    
}
\end{verbatim}
}

\subsection*{Package Installation and Getting Started}

The dpeaqms package is an R package which relies on the rjags package
and the native JAGS library already being installed.  The JAGS library
can be downloaded and installed from
\url{http://www-ice.iarc.fr/~martyn/software/jags/}.  The rjags
package is an R interface to this native JAGS library and can be
installed within R by using the command
\begin{verbatim}
install.packages("rjags")
\end{verbatim}
and then the dpeaqms package can be installed by using the command
\begin{verbatim}
install.packages("dpeaqms", 
                 repos="http://r-forge.r-project.org")
\end{verbatim}
Once the dpeaqms package is installed, an overview vignette can be
accessed by using the command
\begin{verbatim}
vignette("dpeaqms")
\end{verbatim}
A vignette showing the analysis of the simulated dataset described in section
\ref{sec:sim} can be accessed using the command
\begin{verbatim}
vignette("dpeaqms.simulatedDataset")
\end{verbatim}

\subsection*{The MCMC sampling scheme}
The MCMC scheme is a Gibbs sampler which involves simulating
  realisations of the model parameters in turn from their full
  conditional distributions as follows:

\begin{itemize}
\item The log-normalisation ratio for reporter ion $i$ in experiment
  $e$ is simulated using $\kappa_{egi}|\cdot\sim \N(A_{egi},1/B)$
  for $(g,i)\neq (g_e,1)$, where
\[
A_{egi}=\frac{a_\kappa b_\kappa
+\sum_{jk}(y_{egjki}-\alpha_{jk}-\beta_{gj}\gamma_{gj})/\sigma^2}
{b_\kappa+n_\kappa/\sigma^2},
\qquad
B_{ei}=b_\kappa+n_\kappa/\sigma^2,
\]
and $n_\kappa=\sum_jm_j$ is the number of log-intensity measurements
for reporter $i$ in experiment $e$.

\item
The mean log-expression level for MS/MS spectrum~$k$ for (control)
group~1 is simulated using $\alpha_{jk}|\cdot\sim \N(C_{jk},1/D)$, where
\[
C_{jk}=\frac{a_\alpha b_\alpha+\sum_{egi}(y_{egjki}-\kappa_{egi}-\beta_{gj}\gamma_{gj})/\sigma^2}
{b_\alpha+n_\alpha/\sigma^2},
\qquad
D=b_\alpha+n_\alpha/\sigma^2,
\]
and $n_\alpha=\sum_{eg} n_{eg}$ is the number of log-intensity
measurements for each mass spectrum~$k$ assigned to protein~$j$.

\item The parameter $\beta_{gj}$ $(g\neq 1)$ indicating whether or not
  the protein~$j$ is differentially expressed for group $g$ with
  respect to control group~1 is simulated using probabilities
\begin{align*}
\pi(\beta_{gj}=0|\cdot) 
&\propto (1-p_{gj})\,\exp\left\{-\,\frac{1}{2\sigma^2}\sum_{eki}\left(y_{egjki}-\kappa_{egi}-\alpha_{jk}\right)^2\right\},\\
\pi(\beta_{gj}=1|\cdot) 
&\propto p_{gj}\,\exp\left\{-\,\frac{1}{2\sigma^2}\sum_{eki}\left(y_{egjki}-\kappa_{egi}-\alpha_{jk}-\gamma_{gj}\right)^2\right\}.
\end{align*}

\item The probability of differential expression of a protein $j$ in
  group $g$ is simulated using $p_{gj}|\cdot\sim
  \text{Beta}(a_p+\beta_{gj},b_p+1-\beta_{gj})$ for $g\neq 1$.

\item 
The difference in the mean log-expression levels of protein~$j$
between group $g$~$(\neq 1)$ and the control group is simulated using
$\gamma_{gj}|\beta_{gj}=0,\cdot\sim N(a_\gamma,1/b_\gamma)$, its prior
distribution, or $\gamma_{gj}|\beta_{gj}=1,\cdot\sim \N(E_{gj},1/F_{gj})$ as
appropriate, where
\[
E_{gj}=\frac{a_\gamma b_\gamma+\sum_{eki}\left(y_{egjki}-\kappa_{egi}-\alpha_{jk}\right)/\sigma^2}
{b_\gamma+n_{gj}^\gamma/\sigma^2},
\qquad
F_{gj}=b_\gamma+n_{gj}^\gamma/\sigma^2
\]
and $n_{gj}^\gamma=m_j\sum_en_{eg}$ is the total number log-intensities
measurements for samples in group~$g$ for protein~$j$.

\item The error standard deviation is simulated using
  $\sigma^{-2}|\cdot\sim\Ga \left(a_\sigma+n/2,b_\sigma+G/2\right)$
  where
  $G=\sum_{egjki}(y_{egjki}-\kappa_{egi}-\alpha_{jk}-\beta_{gj}\gamma_{gj})^2$
  and $n=En_I\sum_j m_j$ is the total number of log-intensity
  measurements for the isobaric labels.
\end{itemize}
\end{document}